# Generation of System Function Maps in Projection-Based Magnetic Particle Imaging Using Lock-in-Amplifier Model

Kenya Murase[1,2*]

[1]Department of Medical physics and Engineering, Division of Medical Technology and Science, Course of Health Science, Graduate School of Medicine, Osaka University, Suita, Osaka, Japan
[2]Global Center for Medical Engineering and Informatics, Osaka University, Suita, Osaka, Japan
*Corresponding author, email: murase@sahs.med.osaka-u.ac.jp

**Abstract**
We previously developed a system for projection-based magnetic particle imaging (MPI) with a field-free-line (FFL) encoding scheme. In the projection-based MPI, projection data are given by the convolution between the system function in the spatial domain and the line integral of the concentration of magnetic nanoparticles (MNPs) through the FFL. Thus, it is important to estimate the system functions and to investigate the factors affecting them for enhancing the quantitative property of the projection-based MPI. The purpose of this study was to present a method for generating the system function maps in projection-based MPI. In the simulation studies, the MPI signals induced by MNPs in a receiving coil were calculated using a lock-in-amplifier model under the assumption that the magnetization and particle size distribution of MNPs obey the Langevin theory of paramagnetism and a log-normal distribution, respectively. The system function maps were generated by calculating the MPI signals at various distances from the FFL and angles between the axis of the receiving coil and the selection magnetic field. The effects of the particle size of MNPs, the viscosity of the suspending medium, the amplitude of the drive magnetic field, and the gradient strength of the selection magnetic field on the system function were investigated. The spatial distributions of the system functions could be well understood from their maps generated by our method and were largely affected by the parameters described above. Our method will be useful for improved understanding, optimization, and development of projection-based MPI.

## 1. Introduction

Magnetic particle imaging (MPI) was introduced in 2005 for imaging of the spatial distribution of magnetic nanoparticles (MNPs) [1]. MPI shows great potential in terms of sensitivity, spatial resolution, and imaging speed [1-5]. MPI exploits the nonlinear magnetization response of MNPs to detect their presence in an alternating magnetic field called the drive magnetic field. Spatial encoding is accomplished by saturating the magnetization of the MNPs almost everywhere except in the vicinity of a special region called the field-free point (FFP) or field-free line (FFL) using a static magnetic field called the selection magnetic field [1-5].

We previously developed a system for projection-based MPI with an FFL-encoding scheme [2, 3]. In the projection-based MPI, projection data are given by the convolution between the system function in the spatial domain and the line integral of the concentration of MNPs through the FFL [2, 3]. Thus, it is important to estimate the system functions and to investigate the factors affecting them for enhancing the quantitative property of the projection-based MPI.

Recently, we have presented a lock-in-amplifier model for analyzing the behavior of signal harmonics in MPI and reported that the behavior of signal harmonics largely depends on the strength of the drive and selection magnetic fields, the particle size distribution of MNPs, and the parameters in the lock-in amplifier such as the time constant of the low-pass filter [6]. Our previous study [6] also suggested that our lock-in-amplifier model is applicable to the estimation of the system functions in MPI. Then, the purpose of this study was to present a method for generating the system function maps in projection-based MPI using our lock-in-amplifier model [6] and to investigate the factors affecting the system functions using simulation studies.

## 2. Materials and Methods

### 2.1. Voltage Induced by MNPs

Figure 1 illustrates the coordinate system for generating the system function maps in projection-based MPI and the relationship between positions of a receiving coil (dotted circle) and an FFL (bold line). As illustrated in Figure 1, we assumed that the FFL is located at spatial position $\mathbf{r} = (0, y, 0)$ and the axis of the receiving coil points in the z direction, perpendicular to the FFL. We also assumed that the drive magnetic field is applied parallel to the axis of the receiving coil, i.e., in the z direction.

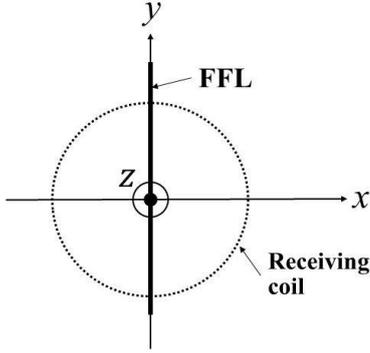

**Figure 1:** Coordinate system for generating the system function maps in projection-based magnetic particle imaging (MPI) and the relationship between positions of a receiving coil (dotted circle) and a field-free line (FFL) (bold line). The axis of the receiving coil points in the z direction, parallel to the drive magnetic field.

When the MNPs with a point-like distribution of concentration $[C_0 \delta(\mathbf{r}), \delta(\mathbf{r})$: Dirac delta function] are located at position $\mathbf{r} [= (x, y, z)]$, and the receiving coil sensitivity is assumed to be constant ($\sigma_0$), the changing magnetization of the MNPs induces a voltage $[v_{rx}(t)]$ according to Faraday's law, which is given by [7]

$$v_{rx}(t) = -\mu_0 \sigma_0 C_0 \frac{\partial M_z(\mathbf{r}, t)}{\partial t} \quad (1)$$

where $M_z(\mathbf{r}, t)$ is the z component of the magnetization of MNPs per unit concentration at position $\mathbf{r}$ and time $t$, and $\mu_0$ is the magnetic permeability of a vacuum. In the following, we neglected constant factors in Equation (1) for generalization. In addition, we assumed that the signal obtained by the receiving coil includes Gaussian white noise [6].

When the relaxation effect of MNPs cannot be neglected, the voltage induced by MNPs $[\tilde{v}_{rx}(t)]$ can be approximated by the temporal convolution of $v_{rx}(t)$ with an exponential kernel [8, 9], i.e.,

$$\tilde{v}_{rx}(t) = v_{rx}(t) \otimes \left[ \frac{1}{\tau_r} e^{-\frac{t}{\tau_r}} \cdot h(t) \right] \quad (2)$$

In Equation (2), $\otimes$ denotes the convolution integral, $h(t)$ is the Heaviside step function, and $\tau_r$ is the effective relaxation time given by $1/\tau_r = 1/\tau_N + 1/\tau_B$, where $\tau_N$ and $\tau_B$ are the Néel relaxation time and Brownian relaxation time, respectively. $\tau_N$ and $\tau_B$ are given by $\tau_N = \tau_0 \sqrt{\pi} e^\Gamma / (2\sqrt{\Gamma})$ and $\tau_B = 3\eta V_H / (k_B T)$ [8, 9], where $\tau_0$ is the average relaxation time in response to a thermal fluctuation, $\eta$ the viscosity of the suspending medium, $k_B$ the Boltzmann constant, $T$ the absolute temperature, and $\Gamma = K V_M / (k_B T)$ with $K$ and $V_M$ being the anisotropy constant of MNPs and the magnetic volume given by $V_M = \pi D^3 / 6$ for a particle of diameter $D$, respectively. $V_H$ is the hydrodynamic volume of MNPs given by $V_H = (1 + 2\delta/D)^3 V_M$, where $\delta$ is the thickness of the surfactant layer [8, 9].

### 2.2. MPI Signal

The magnetization of MNPs per unit concentration at position $\mathbf{r}$ and time $t$ $[M(\mathbf{r}, t)]$ in response to an applied magnetic field can be described by the Langevin function [10] as

$$M(\mathbf{r}, t) = M_s V_M \left( \coth \xi - \frac{1}{\xi} \right) = M_s V_M \mathcal{L}(\xi) \quad (3)$$

where $M_s$ is the saturation magnetization and $\xi$ is the ratio of the magnetic energy of a particle with magnetic moment $m$ in an external magnetic field to the thermal energy given by $k_B T$:

$$\xi = \frac{\mu_0 m H(\mathbf{r}, t)}{k_B T} = \frac{\mu_0 M_d V_M H(\mathbf{r}, t)}{k_B T} \quad (4)$$

In Equation (4), $M_d$ is the domain magnetization of MNPs. $H(\mathbf{r}, t)$ is the strength of the external magnetic field at position $\mathbf{r}$ and time $t$. When the drive magnetic field is applied parallel to the axis of the receiving coil (Figure 1) in the presence of the selection magnetic field, $H(\mathbf{r}, t)$ is given by

$$H(\mathbf{r}, t) = \sqrt{H_S(\mathbf{r})^2 + H_D(t)^2 + 2H_S(\mathbf{r})H_D(t)\cos\phi} \quad (5)$$

where $H_S(\mathbf{r})$, $H_D(t)$, and $\phi$ denote the strength of the selection magnetic field at position $\mathbf{r}$, the strength of the drive magnetic field at time $t$, and the angle between the axis of the receiving coil and the selection magnetic field, respectively. $H_S(\mathbf{r})$ is given by

$$H_S(\mathbf{r}) = \sqrt{(G_x \cdot x)^2 + (G_y \cdot y)^2 + (G_z \cdot z)^2} \quad (6)$$

where $G_x$, $G_y$, and $G_z$ denote the x, y, and z components of the gradient strength of the selection magnetic field, respectively. $\cos\phi$ in Equation (5) is given by

$$\cos\phi = \frac{G_z \cdot z}{H_S(\boldsymbol{r})} \qquad (7)$$

In this study, we assumed that $H_D(t)$ is given by

$$H_D(t) = A_D \cos(2\pi f_D t) \qquad (8)$$

where $A_D$ and $f_D$ denote the amplitude and frequency of the drive magnetic field, respectively.

The z component of $H(\boldsymbol{r}, t)$ is given by

$$H_z(\boldsymbol{r}, t) = H_D(t) + H_S(\boldsymbol{r})\cos\phi \qquad (9)$$

Thus, we obtain

$$\begin{aligned}\frac{\partial M_z(\boldsymbol{r},t)}{\partial t} &= M_s V_M \left\{\beta \frac{d\mathcal{L}(\xi)}{d\xi}\left(\frac{H_z}{H}\right)^2 \right. \\ &\quad \left. - \frac{\mathcal{L}(\xi)}{H}\left[\left(\frac{H_z}{H}\right)^2 - 1\right]\right\}\frac{\partial H_z}{\partial t} \\ &= -2\pi f_D A_D M_s V_M \left\{\beta \frac{d\mathcal{L}(\xi)}{d\xi}\left(\frac{H_z}{H}\right)^2 \right. \\ &\quad \left. - \frac{\mathcal{L}(\xi)}{H}\left[\left(\frac{H_z}{H}\right)^2 - 1\right]\right\}\sin(2\pi f_D t)\end{aligned} \qquad (10)$$

where $\beta = \mu_0 M_d V_M/(k_B T)$, and $\mathcal{L}(\xi)$ and $\xi$ are given by Equation (3) and Equation (4), respectively. Note that $H(\boldsymbol{r},t)$ and $H_z(\boldsymbol{r},t)$ are denoted by $H$ and $H_z$, respectively, for simplicity.

At the instantaneous position $(x, 0, z)$ from the FFL at $(0, 0, -H_D(t)/G_z)$, Equation (10) becomes [11]

$$\frac{\partial M_z(\boldsymbol{r},t)}{\partial t} = M_s V_M \left\{\beta \frac{d\mathcal{L}(\xi)}{d\xi}\left(\frac{G_z \cdot z}{H}\right)^2 + \frac{\mathcal{L}(\xi)}{H}\left(\frac{G_x \cdot x}{H}\right)^2\right\} G_z \cdot v_z \qquad (11)$$

where $H = \sqrt{(G_x \cdot x)^2 + (G_z \cdot z)^2}$ and $v_z$ denotes the velocity of the FFL. The first and second terms in the right hand side of Equation (11) denote the time derivative of $M_z(\boldsymbol{r},t)$ proportional to the tangential ($z$) and normal ($x$) components of the FFL velocity vector, respectively [11]. On the trajectory line of the FFL, *i.e.*, at $x = 0$, the second term in the right hand side of Equation (11) vanishes and then Equation (11) is reduced to

$$\frac{\partial M_z(\boldsymbol{r},t)}{\partial t} = M_s V_M \beta \frac{d\mathcal{L}(\xi)}{d\xi} G_z \cdot v_z \qquad (12)$$

On the line perpendicular to the FFL trajectory, *i.e.*, at $z = 0$, Equation (11) is reduced to

$$\frac{\partial M_z(\boldsymbol{r},t)}{\partial t} = M_s V_M \frac{\mathcal{L}(\xi)}{G_x|x|} G_z \cdot v_z \qquad (13)$$

where $|x|$ denotes the absolute value of $x$.

## 2.3. Particle Size Distribution

We assumed that the particle size distribution obeys a log-normal distribution [12]. Thus, the magnetization of MNPs [$M(\boldsymbol{r},t)$ given by Equation (3)] averaged based on this particle size distribution [$\langle M(\boldsymbol{r},t)\rangle$] is given by [13, 14]

$$\begin{aligned}&\langle M(\boldsymbol{r},t)\rangle \\ &= \frac{1}{\sqrt{2\pi}}\int_0^\infty \frac{M(\boldsymbol{r},t)}{\sigma D}\exp\left[-\frac{1}{2}\left(\frac{\ln(D)-\mu}{\sigma}\right)^2\right]dD\end{aligned} \qquad (14)$$

where $\mu$ and $\sigma$ denote the mean and standard deviation of the log-normal distribution of $D$, respectively [13, 14].

## 2.4. Generation of System Function Maps Using Lock-in-Amplifier Model

As described in our previous paper [6], a lock-in amplifier performs signal mixing, *i.e.*, a multiplication of its input [$\tilde{v}_{rx}(t)$ given by Equation (2)] with a reference signal. Because the third-harmonic signal is used as the MPI signal ($S_{MPI}$) [2, 3], the frequency of the reference signal was taken as the three-fold frequency of the drive magnetic field [$f_D$ in Equation (8)]. The mixed signal is then fed through an adjustable low-pass filter to extract the output signal from the DC component of the filtered signal. In this study, $S_{MPI}$ was defined as the mean absolute value of the output signal [6, 15].

The relationship between $S_{MPI}$ and the distance from the FFL corresponds to the system function in the spatial domain in the projection-based MPI. The maps of the system function were generated by calculating the $S_{MPI}$ values for $x$ and $z$ ranging from -10 mm to 10 mm with steps of 0.25 mm.

## 2.5. Simulation Studies

We considered magnetite (Fe$_3$O$_4$) as MNPs, and $M_d$ in Equation (4) was taken as 446 kA/m [9, 13, 14]. $M_s$ in Equation (3) is given by the product of $M_d$ and the volume fraction of MNPs [13, 14]. In this study, the volume fraction of MNPs was assumed to be 1.0 for simplicity. When estimating $\tau_r$ in Equation (2), the anisotropy constant of MNPs ($K$) and the thickness of the surfactant layer ($\delta$) were assumed to be 9000 J/m$^3$ and 2 nm, respectively [9, 13, 14].

Unless specifically stated, the amplitude and frequency of the drive magnetic field [$A_D$ and $f_D$ in Equation (8), respectively] were fixed at 10 mT and 400 Hz, respectively [2, 3]. The time constant of the low-pass filter used in the lock-in amplifier and the signal-to-noise ratio [6] were fixed at 10 ms and 20, respectively. The mean diameter of MNPs and $\sigma$ in Equation (14) were assumed to be 20 nm and 0.2, respectively, and the viscosity of the suspending medium ($\eta$) was fixed at 0.005 kg/m/s. The temperature ($T$) was assumed to be room temperature (293.15 K). $G_x$, $G_y$, and $G_z$ in Equation (6) were assumed to be 2 T/m, 0 T/m, and 2 T/m, respectively.

When investigating the effect of the mean diameter of MNPs on the system function, the mean diameter of MNPs was varied as 15 nm, 20 nm, and 30 nm. When investigating the effect of the viscosity of the suspending medium, $\eta$ was varied as 0.001 kg/m/s, 0.005 kg/m/s, and 0.01 kg/m/s. In this case, the mean diameter of MNPs was taken as 30 nm. When investigating the effect of the amplitude of the drive magnetic field, $A_D$ was varied as 5 mT, 10 mT, and 15 mT. When investigating the effect of the gradient strength of the selection magnetic field, the combination of $G_x$ and $G_z$ was varied as 2 mT and 1 mT, 1 mT and 2 mT, 2 mT and 2 mT, 4 mT and 4 mT, and 6 mT and 6 mT.

## 3. Results

Figure 2 shows the system function maps (left panels) and horizontal (x axis) and vertical profiles (z axis) through the center of the maps (right panels) for various mean diameters of MNPs [(a) for 15 nm, (b) for 20 nm, and (c) for 30 nm]. As shown in the left panels in Figure 2, the spatial distribution of the system function differed largely in the x and z directions. The system function decreased monotonically with increasing distance from the FFL in the x direction, whereas oscillation including a dent was observed in the z direction. The system function values also changed depending on the mean diameter of MNPs (right panels in Figure 2).

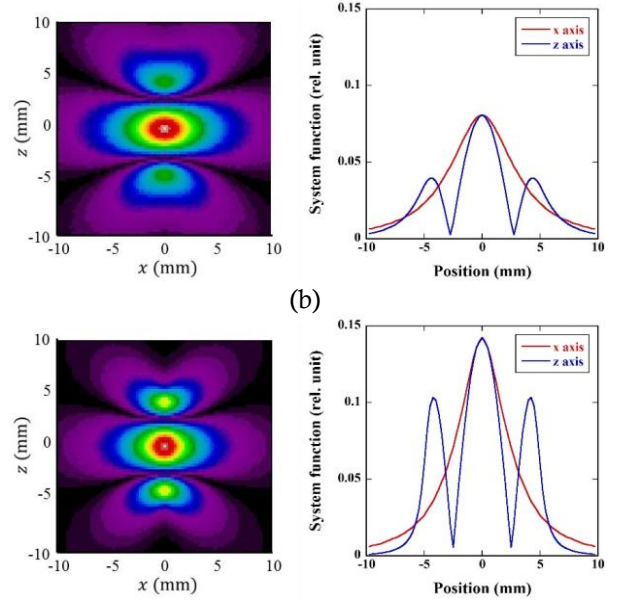

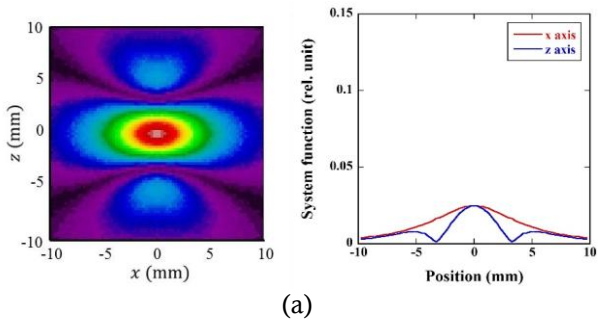
(a)

**Figure 2:** System function maps (left panels) and horizontal (x axis) and vertical profiles (z axis) through the center of the maps (right panels) for various mean diameters of magnetic nanoparticles [(a) for 15 nm, (b) for 20 nm, and (c) for 30 nm]. The vertical scales in the right panels were adjusted to be the same.

Figure 3 shows the system function maps (left panels) and horizontal (x axis) and vertical profiles (z axis) through the center of the maps (right panels) for various $\eta$ values [(a) for 0.001 kg/m/s, (b) for 0.005 kg/m/s, and (c) for 0.01 kg/m/s]. Although the system function maps showed similar spatial distribution (left panels in Figure 3), the system function value decreased with increasing $\eta$ value (right panels in Figure 3).

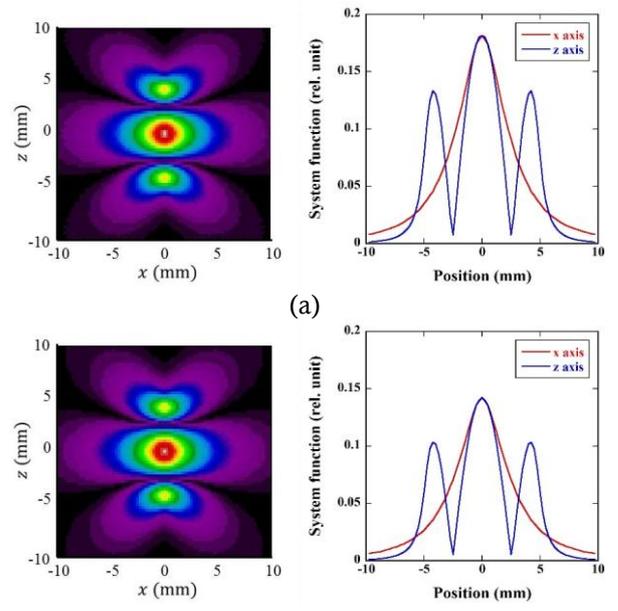

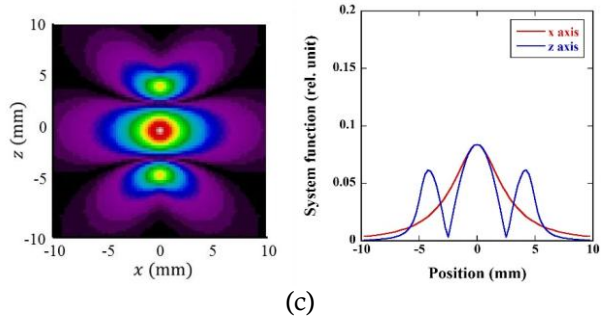

(c)

**Figure 3:** System function maps (left panels) and horizontal (x axis) and vertical profiles (z axis) through the center of the maps (right panels) for various viscosities of the suspending medium [(a) for 0.001 kg/m/s, (b) for 0.005 kg/m/s, and (c) for 0.01 kg/m/s]. The vertical scales in the right panels were adjusted to be the same.

Figure 4 shows the system function maps (left panels) and horizontal (x axis) and vertical profiles (z axis) through the center of the maps (right panels) for various $A_D$ values [(a) for 5 mT, (b) for 10 mT, and (c) for 15 mT]. As shown in the left panels in Figure 4, the spatial distribution of the system function spread with increasing $A_D$ value. This is also confirmed from the profiles of the system function shown in the right panels in Figure 4.

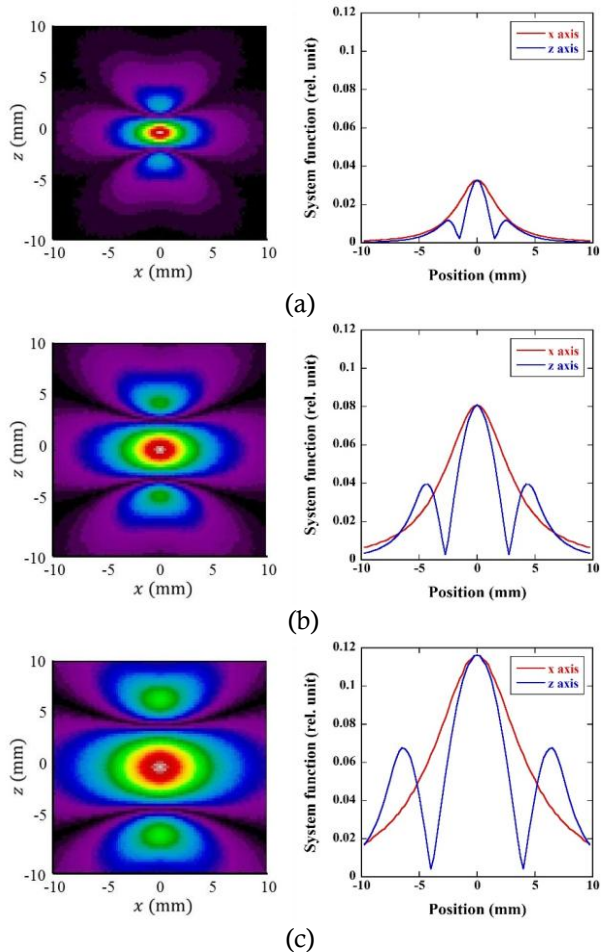

(a)

(b)

(c)

**Figure 4:** System function maps (left panels) and horizontal (x axis) and vertical profiles (z axis) through the center of the maps (right panels) for various amplitudes of the drive magnetic field [$A_D$ in Equation (8)] [(a) for 5 mT, (b) for 10 mT, and (c) for 15 mT]. The vertical scales in the right panels were adjusted to be the same.

Figure 5 shows the system function maps (left panels) and horizontal (x axis) and vertical profiles (z axis) through the center of the maps (right panels) for various combinations of $G_x$ and $G_z$ [(a) for 2 T/m and 1 T/m, (b) for 1 T/m and 2 T/m, (c) for 2 T/m and 2 T/m, (d) for 4 T/m and 4 T/m, and (e) for 6 T/m and 6 T/m]. The spatial distribution of the system function changed depending on $G_x$ and/or $G_z$, *i.e.*, the spatial distribution decreased with increasing $G_x$ and/or $G_z$, and vice versa (left panels in Figure 5). This is also confirmed from the profiles shown in the right panels in Figure 5.

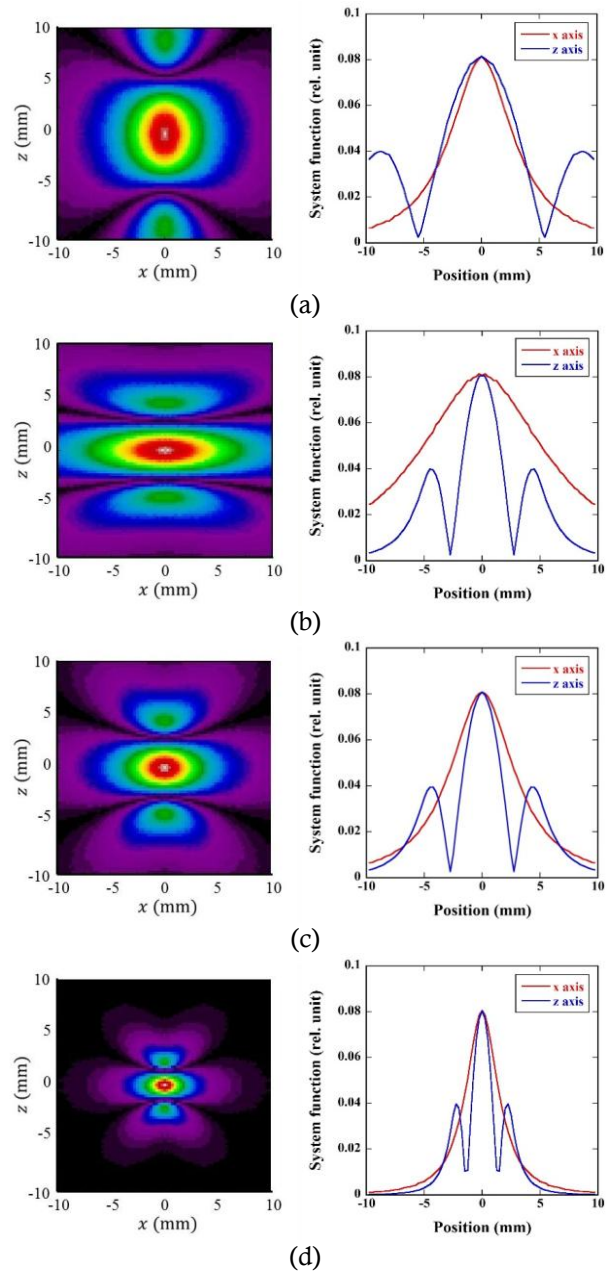

(a)

(b)

(c)

(d)

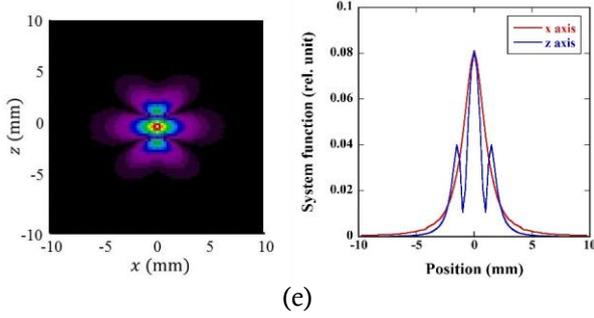
(e)

**Figure 5:** System function maps (left panels) and horizontal (x axis) and vertical profiles (z axis) through the center of the maps (right panels) for various combinations of the gradient strengths of the selection magnetic field in the x ($G_x$) and z directions ($G_z$) [(a) for 2 T/m and 1 T/m, (b) for 1 T/m and 2 T/m, (c) for 2 T/m and 2 T/m, (d) for 4 T/m and 4 T/m, and (e) for 6 T/m and 6 T/m]. The vertical scales in the right panels were adjusted to be the same.

## 4. Discussion

In this study, we presented a method for generating the system function maps in projection-based MPI using the lock-in-amplifier model [6] and investigated the effects of the particle size of MNPs, the viscosity of the suspending medium, the amplitude of the drive magnetic field, and the gradient strength of the selection magnetic field on the system function by simulation studies. Our results (Figures 2-5) demonstrated that our method is useful for visualizing the spatial distribution of the system function in the projection-based MPI, and that the system function in the projection-based MPI is largely affected by the parameters described above.

As shown in the system function maps generated in this study (left panels in Figures 2-5), oscillation including a dent appears in the z direction, whereas this oscillation is not seen in the x direction. As previously described, the axis of the receiving coil points in the z direction, parallel to the drive magnetic field. Thus, the selection magnetic field is parallel to the drive magnetic field on the z axis, *i.e.*, $\phi$ in Equation (5) or Equation (9) is equal to 0° on the z axis, whereas the selection magnetic field is perpendicular to the drive magnetic field on the x axis, *i.e.*, $\phi = 90°$ on the x axis. It is known from Equation (9) that $H_Z(\mathbf{r},t)$ becomes equal to $H_D(t)$ at $\phi = 90°$ and oscillates periodically around $H_Z(\mathbf{r},t) = 0$ irrespective of the selection magnetic field. Furthermore, $H(\mathbf{r},t)$ given by Equation (5) increases with increasing distance from the FFL. Thus, $S_{MPI}$ appears to decrease monotonically with increasing distance from the FFL. We believe this explains why oscillation including a dent is not seen on the x axis. In contrast, $H_Z(\mathbf{r},t)$ becomes equal to $H_D(t) + H_S(\mathbf{r})$ at $\phi = 0°$ and oscillates periodically around $H_Z(\mathbf{r},t) = H_S(\mathbf{r})$. Thus, it appears that $S_{MPI}$ changes depending on the strength of the selection magnetic field and does not decrease monotonically with increasing distance from the FFL. This appears to be the reason why oscillation including a dent is seen on the z axis.

As previously described, the projection data obtained by projection-based MPI [2, 3] are considered to be given by the convolution between the system function in the spatial domain and the line integral of the concentration of MNPs through the FFL. This causes blurring in the MPI images reconstructed from the projection data [16]. Thus, it appears that the quantitative property of the projection-based MPI can be improved by deconvolution of the system function from the projection data [16]. Our method will be useful for enhancing the quantitative property of projection-based MPI.

Recently, multi-color MPI has been introduced, in which the signals generated by different particle types or particles in different environments are assigned to differently colored images using the system functions measured in advance by calibration scans [17]. As shown in Figure 2 and Figure 3, the system functions obtained by our method change depending on the particle size of MNPs and the viscosity of the suspending medium. These results suggest that our method may be useful for realizing the multi-color MPI using projection-based MPI (projection-based multi-color MPI) and for discriminating different particle types or particles in different environments. Such studies are in progress.

As shown in Figure 4 and Figure 5, the system function maps generated by our method are helpful for visually understanding the dependences of the system function on the amplitude of the drive magnetic field and the gradient strength of the selection magnetic field. This will be useful for the optimization and development of projection-based MPI.

## 5. Conclusion

We presented a method for generating the system function maps in projection-based MPI using our lock-in-amplifier model and investigated the factors affecting the system functions by simulation studies. Our method will be useful for improved understanding, optimization, and development of projection-based MPI.

**ACKNOWLEDGEMENT**

This work was supported by a Grant-in-Aid for Scientific Research from the Japan Society for the Promotion of Science (JSPS) and the Japan Science and Technology Agency (JST).